\documentclass[11pt]{article}
\usepackage[T1]{fontenc}
\usepackage[ansinew]{inputenc}
\usepackage{amsmath}
\usepackage{amssymb}
\usepackage{graphicx}
\usepackage{lscape}
\begin{document}
\title{Asymptotic normalization coefficients (nuclear vertex constants) for  $p+^7Be\to ^8B$  and
   the    direct $^7Be(p,\gamma)^8B$ astrophysical
S-factors at solar energies
}
\author{S.B. Igamov and R. Yarmukhamedov \thanks{Corresponding author, E-mail: rakhim@inp.uz}}
\maketitle {\it Institute of Nuclear Physics, Uzbekistan Academy
of Sciences,100214 Tashkent, Uzbekistan}

\begin{abstract}
A new analysis of the precise experimental astrophysical S-factors
for the direct capture $^7Be(p,\gamma)$ $^8B$ reaction
[A.J.Junghans et al.Phys.Rev. C 68 (2003) 065803 and L.T. Baby  et
al. Phys.Rev. C 67 (2003) 065805] is carried out based on    the
modified two - body potential approach in which the direct
astrophysical S-factor, $ {\rm S_{17}(E)}$, is expressed in terms
of the asymptotic normalization constants for  $p+^7Be\to ^8B$ and
two additional conditions are involved to verify the peripheral
character of the reaction under consideration. The Woods-Saxon
potential form is used for the bound ($p+^7Be$)- state wave
function and for the $p^7Be$- scattering wave function. New
estimates are obtained for the ``indirectly measured'', values of
the asymptotic normalization constants ( the nuclear vertex
constants) for the $p+^7Be\to ^8B$ and $S_{17}(E)$ at E$\le$ 115
keV, including $E$=0. These values of $S_{17}(E)$ and asymptotic
normalization constants have been used for getting information
about the ``indirectly measured'', values of the $s$ wave average
scattering length and the $p$ wave effective range parameters for
$p^7Be$- scattering.
\end{abstract}

\maketitle



\section{Introduction}

 The $^7Be(p,\gamma)^8B$ reaction rate given in terms
of the zero-energy astrophysical S-factor ${\rm  S_{17}(0)}$ is
one of the main input data in the solar neutrino problem  because
high energy neutrinos are produced via the decay $^8B\to
^7Be+e^++\nu_e$ [1-3]. This quantity is determined by both
extrapolating the measured absolute cross sections
$\sigma^{exp}(E)$ ( or equivalently its experimental S-factors
$S^{exp}_{17}(E)$) to the astrophysically relevant energies
($\sim$ 20 keV) [2-4] and by theoretical predictions (see, e.g.
Refs.[5-7] for   review and Refs.[8-11]).

Despite  the  steady and impressive progress in our understanding
of this reaction  made in recent years    in measurements  of $S
_{17}^{exp}(E)$ at extremely low energies \cite{Jun03} and the
theoretical predictions  for $S _{17}(E)$ at solar energies
(0$\le$ E$\lesssim$ 25 keV ) \cite{Des04}, some ambiguities
associated with the prediction for $S_{17}(0)$ however still
exist, and   may considerably influence the predictions of the
standard solar model \cite{Bah89,Bah98}.

Experimentally,  there are two types of data for the $^7{\rm
Be}(p,\gamma)^8B$ cross sections at extremely low energies: i)
eight direct measured data using radioactive $^7Be$ targets with
quoted uncertainties   up to 20\% [7,12-19]. All of these measured
data have a similar energy dependence for the astrophysical
S-factors but the extrapolation of  each of the measured data
from the observed energy ranges to low experimentally inaccessible
energy regions, including E=0, gives a value of $S_{17}(0)$ with
an uncertainty exceeding noticeably the experimental  one.  Here
one notes the recent  values of $S_{17}(0)$ recommended in
Ref.\cite{Jun03} and \cite{Baby03} are 21.4$\pm$0.5(exp)$\pm$
0.6(theor) eVb and 21.2$\pm$ 0.7 eVb, respectively, which have
been obtained from the analysis of the precisely measured data for
${\rm S_{17}(E)}$ by means of an artificial renormalization of the
energy dependence of the microscopic cluster-model calculation
\cite{Des94} for $S_{17}(E)$ to the experimental data. ii)
Indirectly measured data [20-27] obtained from the Coulomb breakup
experiments in which a radioactive beam of $^8B$ nuclei is
dissociated into two fragments ( proton ($p$) and $^7Be$) in the
field of multicharged heavy nuclei. The $^7Be(p,\gamma)^8B$
astrophysical S-factors extracted by the authors of those works
change within the range 16.7$\le$ S(0)$\le$ 20.6 eVb. It is  seen
from here that, there is a discrepancy between the results of two
types of the experimental data, the direct and indirect ones, for
$S_{17}(0)$.   The main reasons of this discrepancy is not known
yet. Besides, as noted in \cite{Jun02} the astrophysical S-factor
$S_{17}(0)$ must be known to $\pm$ 5\% in order that its
uncertainty not be the dominant error in prediction of the solar
neutrino flux \cite{Bah98}.

The theoretical calculations of   $S_{17}(0)$ performed within
different methods also show  considerable spread
\cite{Bar95,Des94,Tim97,Des04,Xu94}. However, the microscopic
cluster-model calculations performed in
Refs.\cite{Des94,Tim97,Des04} show, firstly, considerable
sensitivity of $S_{17}(0)$ to the used effective NN-potential and,
secondly, a correlation between the calculated asymptotic
normalization coefficients (ANC)(or the respective nuclear vertex
constants (NVC) \cite{Blok77}) for $p+^7Be\to ^8B$ (for the
virtual decay $^8B\to p+^7Be $) and the calculated $S_{17}(0)$ has
been revealed. Such correlation does happen since at extremely low
energies, due to the strong Coulomb repulsion and rather low
binding energy (0.137 MeV) of $^8B$ in the ($ p+^7Be $)-channel,
the direct radiative capture $^7{\rm Be}(p,\gamma)^8B$ reaction
proceeds mainly in the region well outside the range of the
internuclear $p^7Be$ interaction \cite{Chris61}. In this case,
$S_{17}(0)$ is expressed in terms of the ANC for   $p+^7Be\to ^8B$
\cite{Tim97,Mukh90a,Xu94}. In this connection, one should also
note the value of ${\rm S_{17}(0)}$=18.2$\pm$ 1.8 eVb
\cite{MGT,Tab06} inferred in the ANC-method by using the values of
the ANC's  for $p+^7Be\to ^8B$ which have been obtained from the
analysis results for the peripheral proton transfer  reactions
performed within the modified DWBA approach [33-35].   But,    the
authors of Ref.\cite{Tab06} also noted that the reason of the
discrepancy between the ${\rm S_{17}(0)}$ result obtained in
\cite{Tab06} and that recommended  in Ref.\cite{Jun03} is not
understood yet. Moreover, as it is noted also by authors of
Refs.\cite{Jun02,Jun03,Baby03}, the ANC-method is still subject to
uncertainties related to the model dependence of the ANC and the
extracted $S_{17}(0)$ values (see Refs. \cite{Igam07,Igam072} and
below also). From our point of view one of the possible reasons of
the observed discrepancy between the results of Refs.\cite{Jun03}
and \cite{Tab06} is apparently connected with the fact that the
available values of the ANC's for $p+^7Be\to ^8B$ obtained in
[33-35] may not have enough accuracy \cite{Igam072}. Therefore,
determination of precise experimental values of the ANC's for
$p+^7Be\to ^8B$ is highly desirable since it has direct effects in
the correct extrapolation of the $^7{\rm Be}(p,\gamma)^8B$
astrophysical S-factor at solar energies. For this aim   all
possible applications of the two-body potential model are not
exhausted yet.

In this work    a new analysis of the highly precise experimental
astrophysical S-factors for the direct capture $^7{\rm
Be}(p,\gamma)^8B$ reaction at the energy regions 116$\le$ E$\le$
400 keV and 1000$\le$ E$\le$ 1200 keV \cite{Jun03,Jun02,Baby03} is
performed within the modified two - body potential approach
\cite{Igam07} to obtain ``indirectly measured'', values both of
the ANC's (the NVC's) for $p+^7{\rm Be}\to ^8B$ and of $S_{17}(E)$
at E$\le$ 115 keV, including E=0. In the present work we show that
one can extract ANC's for $p+^7Be\to ^8B$ directly from the
$^7{\rm Be}(p,\gamma)^8B$ reaction where the ambiguities inherent
for the standard two -body potential model calculation of the
$^7{\rm Be}(p,\gamma)^8B$ reaction , connected with the choice of
the geometric parameters (the radius $r_o$ and the diffuseness
$a$) for the Woods-Saxon potential and the spectroscopic factors
(for example, see Refs.[15, 38-43] and below), can be reduced in
the physically acceptable limit, being within the experimental
errors for the $S^{exp}_{17}(E)$ reached in
\cite{Jun03,Jun02,Baby03}.

The contents of this paper are as follows. In Section 2   basic
formulae of the modified two-body potential approach to the direct
radiative capture $p+^7Be\to^8B+\gamma$ reaction  are given. There
the analysis of the precise measured astrophysical S-factors for
the direct radiative capture $^7Be(p,\gamma)^8B$ reaction is given
(Subsections 2.2-2.4). The conclusion is given in Section 3.

\section{Analysis of   $^7Be(p,\gamma)^8B$  reaction.
Result and discussion}

\subsection{Basic formulae}

\hspace{0.6cm}Here we give the   formulae specialized for the
astrophysical S-factor  to the case of the  $^7{\rm
Be}(p,\gamma)^8B$ reaction. Let us write    $l_f$ ($j_f$)  for the
orbital (total) angular momentum of the proton in the nucleus
$^8B(^7Be+p)$, $l_ i$ for the orbital angular momentum of the
relative motion of the colliding particles in the initial state,
$\lambda $ for multipole order of the electromagnetic transition,
$\eta_f$ for the Coulomb parameter for the $^8B$(=$^7{\rm Be}+p$)
bound state and $\mu$ for  the reduced mass of the ($p^7{\rm Be}$)
pair. For the $^7Be(p,\gamma)^8B$ reaction, the value of $l_f$ is
taken to be equal to 1 and the values of $j_f$ are taken to be
equal to 1/2 and 3/2, while $l_i$=0 and 2 for the $E1$-transition
and $l_ i$=1 for the $E2$-transition.

 According to \cite{MGTPHR01,Igam07}, we can write the astrophysical
 S -factor in the form
\begin{equation} S_{17}(E)=(\sum _{j_f}C^2_ {l_fj_f}){\cal
{R}}_{l_f}(E,C_{l_fj_f}^{(sp)}). \label{12}
\end{equation}
Here, $C_ {l_fj_f}$ is   the ANC  for  $p+^7Be\to ^8B$, which
determines    the amplitude of the tail of the nucleus $^8B$ bound
state wave function in the ($^7Be + p$)-channel and is related to
the NVC $G_{l_fj_f}$ for the virtual decay $^8B\rightarrow ^7Be +
p$ and  to   the spectroscopic factor $Z_{l_fj_f}$ for the
($^7Be+p$)-configuration with the quantum numbers $l_f,j_f$ in the
nucleus $^8B$ as \cite{Blok77}
\begin{equation}
G_{l_fj_f}=-i^{l_f+\eta_f}\frac{\sqrt{\pi}}{\mu}C_{l_fj_f}
\label{6}
\end{equation}
and
\begin{equation}
C_{l_fj_f}= Z_{l_fj_f}^{1/2}C^{(sp)}_{l_fj_f},\label{13a}
\end{equation}
respectively,  and
\begin{equation}
{\cal {R}}_{l_f}(E,C^{(sp)}_{l_fj_f})=\frac{
\tilde{S}_{l_fj_f}(E)}{(C^{(sp)}_{l_fj_f})^2}, \label{13}
\end{equation}
where
$\tilde{S}_{l_fj_f}(E)=\sum_{\lambda}\tilde{S}_{l_fj_f\lambda}$ is
the single-particle astrophysical S-factor and $C^{(sp)}_{l_fj_f}$
is the single particle ANC, which determines the amplitude of the
tail of the single-particle wave function of the bound $^8{\rm
B}$($^7Be + p$) state. In (\ref{6}) the factor taking into account
the nucleon's identity \cite{Blok77} is absorbed in the
$C^{(sp)}_{l_fj_f}$. The single-particle bound state wave
function, $\varphi_{l_fj_f}(r)$, is determined by the solution of
the radial Schr\"{o}dinger equation with the phenomenological
Woods - Saxon potential for the given quantum numbers $n$ ($n$ is
the nodes of $\varphi_{l_fj_f}(r)$), $l_f$ and $j_f$ as well as
geometric parameters of $r_o$ and $a$ and with depth adjusted to
fit the binding energy $\epsilon _p$ of the $^8B$ ground state
with respect to the  $^7Be + p$ -channel.
  Note that   in Eq.(\ref{13})  a dependence
of the function ${\cal {R}}_{l_f}(E,C^{(sp)}_{l_fj_f})$ on the
free parameter $C^{(sp)}_{l_fj_f}$ enters also through the
single-particle wave function $\varphi_{l_fj_f}(r; C^{(sp)}_{
l_fj_f})(\equiv\varphi_{l_fj_f}(r))$ \cite{Gon82}, and the
single-particle ANC $C^{(sp)}_{l_fj_f}$ itself in turn is a
function of the geometric parameters of   $r_o$ and  $a$.

 According to \cite{Igam07}, the peripheral character for the
direct capture $^7Be(p,\gamma)^8B$ reaction are conditioned by
\begin{equation} {\cal {R}}_{l_f}(E,C^{(sp)}_{l_fj_f})=f(E)  \label{15}
\end{equation}
as a function of the $C_{l_fj_f}^{(sp)}$   within the energy range
$E_{min}\leq E \leq E_{max}$, where the left hand side  of
Eq.(\ref{15}) must not depend on $C_{l_fj_f}^{(sp)}$ for each
fixed $E$ from the aforesaid energy range, and by
\begin{equation}
C_{l_f}^2=\frac{S(E)}{{\cal{R}}_{l_f}(E,C_{l_fj_f}^{(sp)})} =const
\label{16}
\end{equation}
for each fixed  $E$ and the function of ${\cal
{R}}_{l_f}(E,C_{l_fj_f}^{(sp)})$  from (\ref{15}), where
$C^2_{l_f}=\sum_{j_f}C^2_{l_fj_f}$.

As  it was previously shown  in \cite{Igam07} for the analysis of
the highly precise experimental astrophysical S-factors
\cite{Brun94}
 for the direct capture $t(\alpha,\gamma )^7 Li $
reaction, fulfillment  of the conditions (\ref{15}) and (\ref{16})
enables one also to obtain valuable information about the
experimental value of the ANC $(C_{l_f}^{exp})^2$ for $^7Be+p
\rightarrow ^8B$ by using $S^{exp}_{17}(E)$ instead of ${\rm
S_{17}(E)}$ in the right hand side (r.h.s.) of Eq.(\ref{16}):
\begin{equation}
(C_{{l_f}}^{exp})^2=\frac{S^{exp}_{17}(E)} {{\cal
{R}}_{l_f}(E,C_{l_fj_f}^{(sp)})}.
  \label{17} \end{equation}
Then the value of the
 ANC, $(C_{l_f}^{exp})^2$, obtained from Eq.(\ref{17})  together with the condition (\ref{15}) can be
used for calculation of $S_{17}(E)$ at    energies of $E<E_{min}$
by the expression:
\begin{equation} S_{17}(E)=(C_{l_f}^{exp})^2{\cal {R}}_{l_f}(E,C^{(sp)}_{l_fj_f}).
 \label{18}
\end{equation}
Values obtained in such a way for the  $(C_{l_f}^{exp})^2$ and
${\rm S_{17}(E)}$ at   energies of $E<E_{min}$ can be considered
as an ``indirect measurement'', of the ANC (or NVC) for the
$^7Be+p\rightarrow ^8B$ and of the astrophysical S-factor for the
direct capture $^7Be(p,\gamma)^8B$ reaction at $E<E_{min}$,
including $E=0$. It should be noted the expressions
(\ref{12})-(\ref{18})   allow  one to determine both the absolute
value of ANC (or NVC) for
 $^7Be+p \rightarrow ^8B$ and that of the astrophysical S-factor $S_{17}(E)$
for the peripheral direct capture $^7Be(p,\gamma)^8B$ reaction at
extremely low experimentally inaccessible  energy regions by means
of the analysis of the same precisely measured values of the
experimental astrophysical S-factor, $S^{exp}_{17}(E)$.

It should be noted that a  condition, similar to the condition
(\ref{15}), has been formulated earlier in [47-49] for the
peripheral character of nucleon (N) transfer reactions within the
modified DWBA approach to determine the ANC for $N+A\to B$.
However, the modified DWBA approach proposed in \cite{Nie,Ar96}
and \cite{Mukh97} and the condition   (see Eq.(9) of
Ref.\cite{Nie} or Eq.(4) of Ref.\cite{Ar96} for example) are
restricted only by the zeroth- and first-other perturbation
approach in the optical Coulomb polarization  potential
$\bigtriangleup V_f^C$ (or $\bigtriangleup V_i^C$) in the
transition operator assuming that the contribution of
$\bigtriangleup V_{f,i}^C$ is  negligible    or  small ,
respectively. In reality, as has been shown in [50-52,37], the
contribution of $\bigtriangleup V_{f,i}^C$ is not small for the
peripheral proton transfer reactions being of astrophysical
interest. Therefore an inclusion of orders (the second and higher)
of the power expansion in $\bigtriangleup V_{f,i}^C$ series is
required in the transition operator of the modified DWBA approach
because they can noticeably influence the absolute normalization
of the peripheral partial amplitudes,  giving the essential
contribution to the DWBA cross-section   calculated near the main
peak \cite{ Yar97,Igam072}. Consequently,  the account of these
expansion terms in the modified DWBA calculations may change
noticeably both the absolute value and the energy dependence of
the left hand side of the condition (9)((4)) of
Ref.\cite{Nie}(\cite{Ar96}), especially, when the proton is
transferred to weakly bound states of the residual nucleus B being
of astrophysical interest \cite{Ar02,Igam072}. Perhaps this is one
of the  main reasons of the fact that the value of ANC (NVC) for
$p+^{16}O\to^{17}F(0.49 MeV)$ extracted in \cite{Ar96,Gag99} from
the analysis of the peripheral $^{16}O(^3He,d)^{17}F$ reaction at
different incident $^3He$-ion energies has an uncertainty up to
about 20\%. The analogous situation takes place for the other
extracted ANC's  of astrophysical interest (see Table 2 in
Ref.\cite{Ar96}). So,   in [47-49], the condition formulated for
the peripheral character for the   transfer reactions is still
subject to uncertainties related to the model dependence of the
extracted ANC values being of astrophysical interest. As for the
conditions (\ref{15}) and (\ref{16}), firstly, they contain only
the parameter $C^{(sp)}_{l_fj_f}$ and, secondly, there the
electromagnetic-transition operator is well known and   taken in
the form of the two-body long wavelength approximation. So the
conditions (\ref{15}) and (\ref{16}), firstly, can be considered
as the necessary and sufficient conditions for the direct capture
$^7Be(p,\gamma )^8B$ reaction   being mainly peripheral and,
secondly, could be used as a test of the reliability of the
modified DWBA calculations [33-35,48,53] for determination of the
ANC-values of astrophysical interest, including those for $^7{\rm
Be}+p\rightarrow ^8B$. These issues for the $^7{\rm
Be}(p,\gamma)^8B$ reaction are considered below.

\subsection{The asymptotic normalization coefficients (nuclear vertex constants)\\ for
 $p+^7Be \rightarrow ^8B$}

\hspace{0.6cm} In this subsection, to determine the ANC (NVC)
values for  $ p +^7Be\to^8B$   the experimental astrophysical
S-factors, $S^{exp}_{17}(E)$, for the $ ^7Be(p,\gamma)^8B$
reaction is reanalyzed      based on    relation (\ref{12}) and
the conditions (\ref{15}) and (\ref{16}) as well as on the
relations (\ref{17}) and (\ref{18}).   The experimental data have
been obtained by authors of [13-16] with experimental uncertainty
being about (or more than) 10\%. Recently, A.R. Junghans et al
\cite{Jun03,Jun02} and L.T. Baby et al \cite{Baby03} apparently
performed the most accurate direct measurement of the $ ^7{\rm
Be}(p,\gamma)^8B$ astrophysical S-factor, covering    the energies
E=116 -2460 keV and 302- 1078 keV with uncertainties 5\% and less
than 10\%, respectively. So, in our analysis we naturally use
$S^{exp}_{17}(E)$ measured in \cite{Jun03,Jun02,Baby03} at
energies 116$\le$ E$\le$ 400 keV  and 1000$\lesssim$ E$\lesssim$
1200 keV, since the reaction under consideration is nonresonant
\cite{BM00,Baby03} and, consequently, proceeds mainly in regions
well outside the range of the internuclear interactions
\cite{Chris61,Des94}. But, in Refs.\cite{Jun03,Jun02}, there are
three   sets (BE1, BE3 S and BE3 L) of experimental data for the
aforesaid energy ranges, which correspond  to different $^8B$
activity (BE1 and BE3) and additional non-common-mode
uncertainties (BE3 S and BE3 L)\cite{Jun03}. Below, we will
separately analyze each of the aforesaid experimental data.

The Woods-Saxon potential [38-43] is used here for the calculation
of both the bound state radial wave function $\varphi _{
l_Bj_B}(r_{p^7Be})$ and  the scattering wave function
$\psi_{l_{p^7Be}j_{p^7Be}}(r_{p^7Be})$ since this potential
reproduces well   the positions of low-lying states of $^8Li$ and
$^8B$ as well as the $s$ wave $n^7Li$-scattering length for the
$I$=2 spin channel and the low energy cross section for the
$n^7Li$-scattering. For the final state, their depth is slightly
modified to fit the binding energy of $^8B$.

The test of the peripheral character of the $^7{\rm
Be}(p,\gamma)^8B$ reaction for the energy ranges of 116$\le$
E$\le$ 400 keV  and 1000$\lesssim$ E$\lesssim$ 1200 keV has been
made by means of verifying the conditions (\ref{15}) and
(\ref{16}), as it was  done in Ref. \cite{Igam07} for the
peripheral $t(\alpha,\gamma)^7Li$ reaction,  by changing the
geometric parameters (radius $r_o$ and diffuseness $a$) of the
adopted Woods-Saxon potential using the procedure of the depth
adjusted to fit the binding energies.

The calculation shows that for each energy E the lion's share of
dependence of the function $\tilde{S}(E)$  on the parameters $r_o$
and $a$ enters mainly through the single-particle ANC
$C^{(sp)}(=C^{(sp)}(r_o,a))$\footnote {Hereafter for simplicity
all indices specifying the singe-particle ANC $C^{(sp)}_{l_f}$ and
the functions ${{\cal {R}}}_{l_f}(E,C^{(sp)}_{l_fj_f})$ and
$\tilde{S}_{l_fj_f}(E)$,  as well as  the induce of the quantum
number $l_f$   specifying the ANC's, NVC's and spectroscopic
factors, have been omitted.}. It should be noted that if one
varies only one parameter $r_o$ or $a$ fixing the other, then
$C^{(sp)}$ changes strongly. But if one varies $r_o$ and $a$ with
the condition $C^{(sp)}=C^{(sp)}(r_o,a)=const$, then the extremely
weak dependence of the $\tilde{S}(E)$ function  on $r_o$ and $a$
for each $C^{(sp)}(r_o,a)=const$ (the ``residual'', ($r_o,
a$)-dependence \cite{Gon82,Igam07}) also occurs. We vary $r_o$ and
$a$ in the wide ranges   ($r_o$ in 0.90-1.50 fm and $a$ in
0.52-0.76 fm) with respect to the standard values ($r_o$=1.20 fm
and $a$=0.65 fm \cite{Bar80,Bar83}). Over this full range,
fulfillment of the conditions (\ref{15}) and (\ref{16}) in the
energy ranges of E$\lesssim$ 400 keV and $1000\le E\le 1200$ keV
is supplied within only $\pm$ 1.1\%. For example, the
``residual'', $(r_o,a)$-dependence  of the single-particle
astrophysical S-factor, $\tilde{S}(E)$, on $r_o$ and $a$ for each
$C^{(sp)}(r_o,a)=const$ turns out to be extremely weak up to about
$\pm$ 1.0\%.  So, $\tilde {S}(E)$ is a rapidly varying function of
$C^{(sp)}$ with the extremely weak ``residual'', ($r_o,a$)-
dependence  for each $C^{(sp)}=const$. However, for each fixed
experimental point of energy E a quantity ${\cal R}(E,C^{(sp)})$
depends  weakly (up to $\pm$ 1.1\%) on the variation of $C^{(sp)}$
, and its ``residual'', ($r_o,a$)-dependence on $r_o$ and  $a$ for
each $C^{(sp)}=const$   is also extremely weak (up to  $\pm$
1.0\%).

\begin{figure}[!h]
\includegraphics{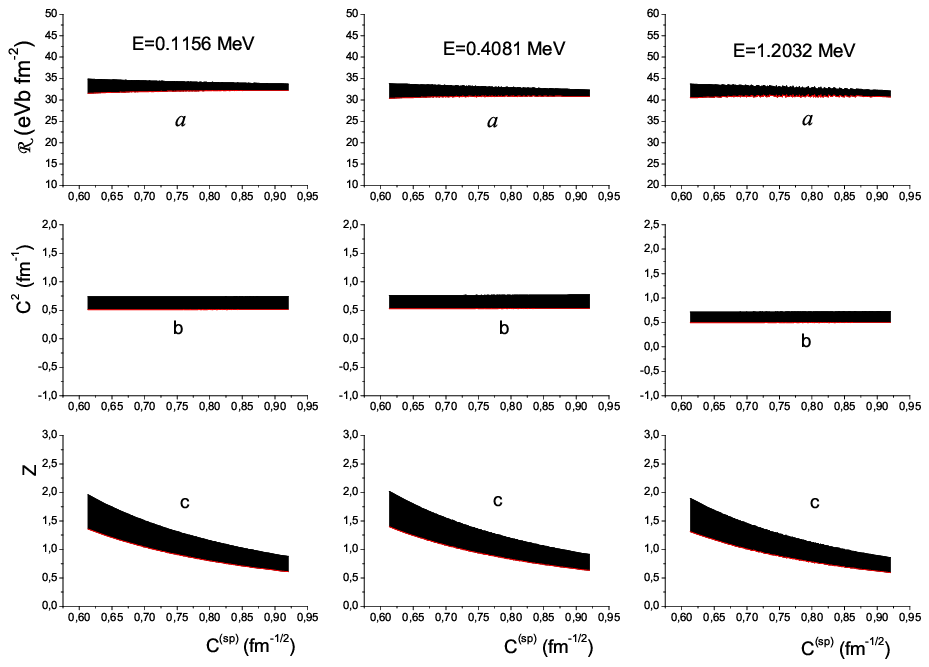}
\caption{\label{fig1}The dependence of
${\cal R}(E,C^{(sp)})$ for the $^7Be(p,\gamma)^8B$ reaction $(a)$,
the ANC $C^2$ for $p+^7Be\to^8B$ $(b)$ and the spectroscopic
factor $Z$ for $^8{\rm B}$ in the $(p+^7Be)$- configuration $(c)$
as a function of the single-particle ANC  $C^{(sp)}$     at
different energies E. The bands show the ranges of uncertainties,
as described in the text.}
\end{figure}

 As an illustration,  Fig.\ref{fig1}$a$  shows  plots of the dependence
of   ${{\cal {R}}}(E,C^{(sp)})$ on the single-particle ANC,
$C^{(sp)}$, only for the three energies $E$. The width of each
band for the curves is the result of the weak ``residual'',
$(r_o,a)$ - dependence of    ${{\cal {R}}}(E,C^{(sp)})$  on the
parameters $r_o$ and $a$ (up to $\pm$ 1.0\% ) for the
$C^{(sp)}=C^{(sp)}(r_o,a)=const$. The same dependence is observed
at other considered energies E. It is seen that for the calculated
values of   ${\cal R}(E,C^{(sp)})$ the dependence on the
$C^{(sp)}$ -value is rather weak (no more than $\pm$ 1.1\%) in the
interval of 0.613$\leq C^{(sp)}\leq$ 0.920 fm$^{-1/2}$ for $
^7{\rm Be}(p,\gamma)^8B$ reaction. It follows from here that the
condition (\ref{15}) is satisfied for the considered reaction
within the uncertainties less than the experimental errors of
$S^{exp}_{17}(E)$.

\begin{figure}[!h]
\includegraphics{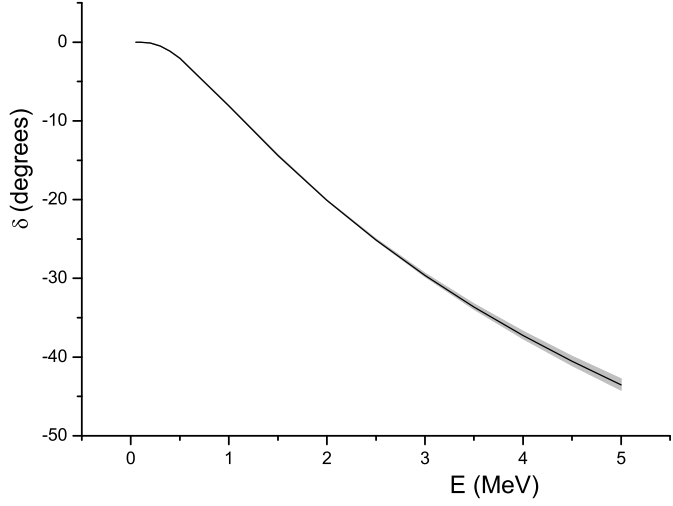} 
\caption{\label{fig2}The energy dependence
of the calculated   $p^7Be$-elastic scattering phase
 shifts for the $s$ wave and the spin channel I=2.   The band is our
calculated data.  The width of the band for fixed energies
corresponds to the variation of the parameters $r_o$ and $a$ of
the adopted Woods - Saxon potential   within the intervals of
$r_o$=0.90 to 1.50 fm and $a$=0.52 to  0.76 fm.}
\end{figure}

We also calculated  the $p^7Be$ - elastic scattering phase shifts
by variation of the parameters $r_o$ and $a$ in the same range for
the adopted Woods-Saxon potential. The calculation performed for
the $s$- wave and the $I$=2 spin channel shows that the $p^7Be$ -
elastic scattering phase shifts change up to 2\% with respect to a
variation of values of the parameters $r_o$ and $a$ in the energy
range of $0\le E\le 5.0$ MeV (see Fig.\ref{fig2}).

Therefore, we test the condition (\ref{16}), which is also
essential for the peripheral character of the reaction under
consideration. For the same energies E as in Fig.\ref{fig1}$a$ we
present in Fig.\ref{fig1}$b$ the results of calculation of the
quantity of
\begin{equation}
C ^2=\frac{S_{17}(E)}{{\cal R}(E,C^({sp)})}, \label{19}
\end{equation}
where instead of   $S_{17}(E)$ the  experimental S- factors,
$S^{exp}_{17}(E)$, for the $ ^7Be(p,\gamma)^8B$ reaction were
taken. Here $C ^2= C ^2_{11/2}+C
^2_{13/2}=C^2_{p_{3/2}}(1+\lambda_C)$, where $C^2_{p_j}=C
^2_{p^7Be;1j}$ and $\lambda_C=C^2_{p_{1/2}}/C^2_{p_{3/2}}$. It is
also noted that the same dependence occurs for other considered
energies. As is seen from   the figure, the obtained values of the
$C^2$ are not practically dependent on the  $C^{(sp)}$-value,
which corresponds to the parameters of the adopted Woods-Saxon
potential $r_o$ ranging within
 0.90-1.50 fm and $a$ in the range of 0.52-0.76 fm (0.613$\leq C^{(sp)}\leq$ 0.920 fm$^{-1/2}$).
However, the values of the spectroscopic factors for $^8B$ in the
$(p+^7Be)$-configuration, $Z$
(=$Z_{11/2}+Z_{13/2}=Z_{p_{3/2}}(1+\lambda_Z)$, where
$Z_{p_j}=Z_{1j}$ and $\lambda_Z=Z_{p_{1/2}}/Z_{p_{3/2}}$),
determined from  Eq.(\ref{13a}) change strongly ( see,
Fig.\ref{fig1}$c$).

For  illustration  Table \ref{table1}  shows  the dependence of
$C^{(sp)}$, $C^2$, $Z$, the single-particle astrophysical
S-factors, $\tilde{S}(E)$, and the ${{\cal {R}}}(E,C^{(sp)})$
functions on the parameters of $r_o$ and $a$ in the aforesaid
regions  for  the reactions under consideration at three energies
$E$.  As  is seen from Table \ref{table1} the uncertainty in
${{\cal {R}}}(E,C^{(sp)})$ , $C^2$ is up to $\pm$ 1\% relative to
the central values of ${{\cal {R}}}(E,C^{(sp)})$ and $C^2$,
obtained for the standard values of $r_o=1.20$ $fm$ and $a=0.65$
$fm$,  for the ($r_o$, $a$) - pair varying in the above mentioned
intervals for $r_o$ and $a$, while the uncertainty in the $Z$   is
about $\pm$ 57\%. Thus, the peripheral character of the reactions
under consideration allows one to determine the $C^2$ value for
$p+^7{\rm Be}\rightarrow ^8B$   with a maximal uncertainty not
exceeding the experimental one for the $S^{exp}_{17}(E)$ when the
geometric parameters $r_o$ and $a$ are varied within the aforesaid
ranges in respect to the standard values of $r_o$ and $a$ and the
precise experimental data from the energy regions of 116$\le$
E$\le$ 400 keV and 1000$\lesssim$ E$\lesssim$ 1200 keV for an
analysis are used.

For different energies E, we also estimate a quantity of a
relative contribution of the nuclear interior ($r\le r_N$) to the
astrophysical S-factors for the  $^7Be(p,\gamma)^8B$ reaction in
dependence on the variation  $C^{(sp)}$ (or $r_o$ and $a$)
introducing  the cutoff radius $r_{cut}$ ($r_{cut}\approx r_N$) in
the lower limit of integration of the radial integral given by
Eq.(10) of Ref.\cite{Igam07}. With this aim one considers the
ratio $ \Delta(E,C^{(sp)};r_{cut})= \mid {\cal
R}(E,C^{(sp)})-\tilde {{\cal R}}(E,C^{(sp)};r_{cut})\mid / {\cal
R}(E,C^{(sp)})$, where $\tilde {\cal R}(E,C^{(sp)};r_{cut})$ is
given by Eqs.(10) of Ref.\cite{Igam07} and (\ref{13}) but in the
radial integral   the integration over $r$ is performed in the
interval $r_{cut}\le r\le\infty$, i.e. $\tilde {{\cal
R}}(E,C^{(sp)};0)={\cal R}(E,C^{(sp)})$. The ${\cal
R}(E,C^{(sp)})$ and  $\tilde {{\cal R}}(E,C^{(sp)};r_{cut})$
functions were calculated for different values of the
single-particle ANC $C^{(sp)}$. A  value of the cutoff radius is
taken  equal to $r_{cut}= r_N= 1.36(7^{1/3}+1)$
\cite{Rolfs79}=3.96 fm,  as well as $r_{cut}$=3.75 fm and 4.25 fm.
The calculation of $\Delta (E,C^{(sp)};r_{cut})$ performed for
different energies $E$ and the aforesaid values of $r_{cut}$ shows
that the quantities of $\Delta(E,C^{(sp)};r_{cut})$  change from
0.1\% up to 2.1\% under variation of $C^{(sp)}$ and
$r_{cut}$\footnote{By using   the case it is worth noticing  here
that there are misprints in \cite{Igam07}. There the left hand
side of Eq.(25) ($\kappa^{2l+1}$) should be replaced by
$(-1)^l\kappa^{2l+1}$ and the phrase ``becomes unambigous'' in the
line 14 upper of page 265 must be written as ``becomes
ambigous'',.}.

Thus the scrupulous analysis performed   here quantitatively
confirms also the conclusion made by different authors (see, for
example Refs.[29-33]) about the fact that the $^7{\rm
Be}(p,\gamma)^8B$ reaction within the considered energy ranges
($E\le$ 400 keV and 1000$\le E\le$ 1200 keV) is strongly
peripheral.

\begin{figure}[!h]
\includegraphics{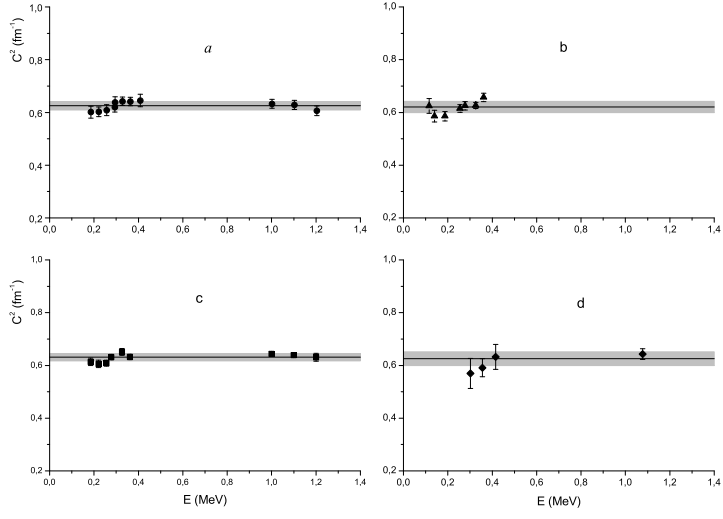} 
\caption{\label{fig3}The  values of the
ANC's, $C^2$,  for  $^7{\rm Be} +p\to^8B$ at all experimental
energies $E $ obtained from data of Refs.\cite{Jun03} (BE1$(a)$,
BE3 L$(b)$ and BE3 S $(c)$) and \cite{Baby03} $(d)$. The solid
lines and the width of the bands
 are the   weighted mean and the   weighted uncertainty, respectively.}
\end{figure}

The experimental values of the ANC's for the  $p +^7{\rm
Be}\rightarrow ^8B$, $ (C^{exp})^2$, are obtained by using the
experimental astrophysical S-factors \cite{Jun03,Jun02,Baby03} in
the r.h.s.  of the relation (\ref{17}) instead of the ${\rm
S_{17}(E)}$ and the central values of ${\cal R}(E,C^{(sp)})$
corresponding to the standard values of the parameters $r_{0}$ and
$a$. The results of the ANC's, $(C^{exp})^2$   obtained separately
for each the aforesaid experimental astrophysical S-factors
and corresponding experimental points of energy $E$ (116$\le$
E$\le$ 400 keV  and 1000$\lesssim$ E$\lesssim$ 1200 keV), are
displayed in Fig.\ref{fig3}. The uncertainty pointed in this
figure for each of the experimental points at energy E corresponds
to that found from (\ref{19}) (averaged square errors (a.s.e.)),
which includes both the absolute  experimental errors in the
corresponding experimental astrophysical S-factor and the
aforesaid uncertainty in ${\cal R}(E,C^{(sp)})$. It is seen from
Fig.\ref{fig3} that the ratio in the r.h.s. of the relation
(\ref{17}) practically does not depend on the energy $E$  although
absolute values of the corresponding experimental astrophysical
S-factors for the reactions under consideration  depend noticeably
on the energy and change by up to 1.33 times in changing E within
the aforesaid energy ranges.

This fact allows us to conclude that the energy dependence of the
experimental astrophysical S-factors \cite{Jun03,Jun02,Baby03} is
well determined by the calculated model-independent function
${\cal R}(E,C^{(sp)})$ and, hence the corresponding experimental
astrophysical S-factors can be used as an independent source of
getting  information about the ANC  for   $p +^7Be\rightarrow
^8{\rm B}$, which determines an absolute normalization of the
direct astrophysical S-factor of the  reaction under consideration
in the aforesaid energy regions, including energies $E$ down up to
zero.

\begin{figure}[!h]
\includegraphics{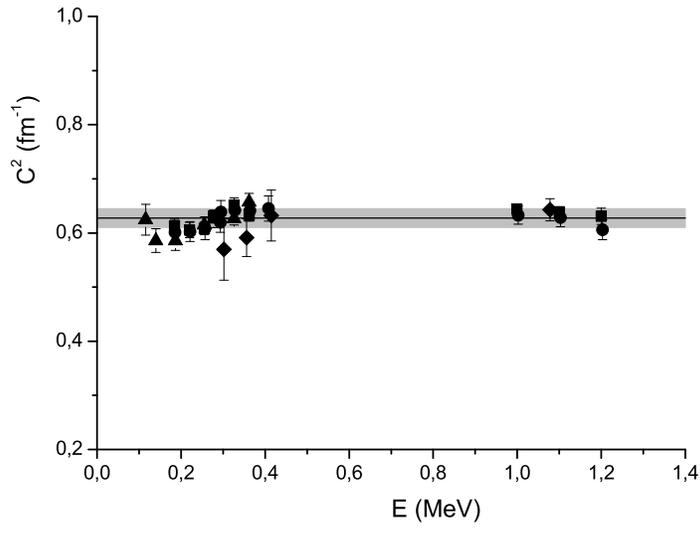}
\caption{\label{fig4}The weighted mean of
$(C^{exp})^2$ for   $p +^7Be\rightarrow ^8B$ (the solid line),
obtained from all the experimental data of Fig.\ref{fig2}. The
points and  the width of the band  are the same data as in
Fig.\ref{fig2}.}
\end{figure}

The weighted means of the ANC - values  ($C^{exp})^2$ for   $p
+^7Be\rightarrow ^8B$, deduced separately from each of the
experimental data, are displayed  by the solid lines in
Fig.\ref{fig3} and  also presented in   the second column of Table
\ref{table2}. The corresponding values of NVC's, $\mid
G\mid_{exp}^2$, are presented in the second-fifth lines of the
third column of Table \ref{table2}. It is seen from Table
\ref{table2} that  the weighted means of the ANC - uncertainty
does not exceed about 4\%. Besides, in Fig.\ref{fig4} (the solid
line) , the weighted mean of  $(C^{exp})^2$ for   $p
+^7Be\rightarrow ^8{\rm B}$, recommended by us and derived from
all the experimental points of Fig.\ref{fig3}, is also presented.
It is equal to $(C^{exp})^2$=0.628$\pm$ 0.017 fm$^{-1}$, which has
the weighted mean uncertainty 2.7\%, and is also in an excellent
agreement with each of the values presented in Table \ref{table2}.
The corresponding value of NVC's is equal to $\mid
G\mid_{exp}^2$=0.114$\pm$ 0.003 fm. This result  for the ANC (NVC)
is the central result of this paper. The latter differs noticeably
from the values of $(C^{exp})^2$=0.462$\pm$0.072 fm$^{-1}$ ($\mid
G \mid_{exp}^2$=0.0835$\pm$ 0.0131 fm) \cite{AMukh1} and
$(C^{exp})^2$=0.466$\pm$ 0.049 fm$^{-1}$ ($\mid
G\mid_{exp}^2$=0.0778$\pm$ 0.0091 fm ) \cite{AMukh2,Tab06}, which
have been respectively obtained from the analysis of the
peripheral proton transfer $^{10}B(^7Be,^8B)^9Be$ and
$^{14}N(^7{\rm Be},^8{\rm B})^{13}C$ reactions performed within
the modified DWBA approach by assuming $\lambda_C$ =0.157
\cite{Tim98,Mukh90} and 0.125 \cite{Tab06,Tra03}, respectively,
and are considered by authors of Ref.\cite{MGTPHR01} as the ``best
value'', determined in a straightforward manner.  But, as it was
mentioned above, the values of the ANC's for $p+^7Be\to ^8B$
obtained in [33-35] may not have the necessary accuracy because
the used modified DWBA approach is the first order perturbation
approximation over the Coulomb polarization operator $\Delta V^C$
in which the latter is assumed to be small \cite{Mukh97}. However,
as shown in Ref.\cite{Igam072} (see also references there), this
assumption is not guaranteed for the aforesaid peripheral proton
transfer reactions since, as it was mentioned above,  an inclusion
of all orders (the second and higher orders) of the power
expansion in a series over $\Delta V^C$ is required in the
transition operator of the DWBA cross section calculations
\cite{AMukh3,Yar97}. Therefore, in reality the values of ANC's
($C^2_{p_{3/2}}$ and $C^2_{p_{1/2}}$) obtained in Refs.[33-35]
must contain additional uncertainties associated with the
aforesaid approximation used in the modified DWBA approach.
Besides, the result obtained by us for the ANC (NVC) also differs
noticeably from the value of $C^2$=0.450$\pm$ 0.039 fm$^{-1}$
($\mid G\mid^2$=0.0816$\pm$ 0.0071 fm) \cite{Trache01} extracted
from data on different breakup reactions at energies of 30-300
MeV/A and that of $(C^{exp})^2$=0.491 fm$^{-1}$ ($ \mid
G\mid_{exp}^2$=0.089 fm) \cite{Bar95}. Since the value of
$\lambda_C$=0.125 is considered as the experimental one
\cite{Tab06,Tra03}, use of the values $(C^{exp})^2$=0.628$\pm$
0.017 fm$^{-1}$ and $\lambda_C$=0.125 leads to the experimental
ANC's (NVC's) values of $(C^{exp}_{p_j})^2$($\mid
G_{p_j}\mid_{exp}^2$), which are equal to
$(C^{exp}_{p_{3/2}})^2$=0.558$\pm$ 0.015 fm$^{-1}$ and
$(C^{exp}_{p_{1/2}})^2$=0.070$\pm$ 0.002  fm$^{-1}$ ($\mid
G_{p_{3/2}}\mid_{exp}^2$=0.0995$\pm$ 0.0027 fm and $ \mid
G_{p_{1/2}}\mid_{exp}^2$=0.0145$\pm$ 0.0004 fm).

  We also would like to note  the results of calculation of ANC's
(NVC's) for   $p +^7Be\rightarrow ^8B$, which were obtained in
Refs.\cite{Tim98,Des04} within the microscopic methods using the
modified Hasegawa-Nagata (MHN), Volkov (V2), M3YE (in
\cite{Tim98}) and Minnesota (MN) (in \cite{Tim98,Des04}) forms of
the NN potential. They are $C^2$=0.975 fm$^{-1}$ ($\mid
G\mid^2$=0.177 fm and $\lambda_C$=0.172)\cite{Tim98} for the MHN
potential, $C^2$=1.157 fm$^{-1}$ ($\mid G\mid^2$=0.210 fm and
$\lambda_C$=0.105) \cite{Tim98} for the V2 potential, $C^2$=0.446
fm$^{-1}$ ($\mid G\mid^2$=0.081 fm and $\lambda_C$=0.157)
\cite{Tim98} for the M3YE potential, $C^2$=0.849 fm$^{-1}$ ($\mid
G\mid^2$=0.154 fm  and $\lambda_C$=0.0405)\cite{Tim98},
$C^2$=0.668 fm$^{-1}$ ($\mid G\mid^2$=0.121 fm and
$\lambda_C$=0.146) and $C^2$=0.770 fm$^{-1}$ ($\mid G\mid^2$=0.140
fm and $\lambda_C$=0.131)\cite{Des04} for the MN- and V2-
potentials, respectively. Besides, one notes the value of
$C^2$=0.556 fm$^{-1}$ ($\mid G\mid^2$=0.101 fm) obtained in
\cite{Des94,Ang03}  within the the microscopic cluster method too.
 It is seen from here that the calculated ANC's $C^2$ and the ratio
$\lambda_C$ are sensitive to the form of the NN potential, and the
values of $C^2$ differ noticeably also from our result. But among
these values of the ANC's, the closest values above and below of
our ANC-value are $C^2$=0.668 fm$^{-1}$ \cite{Des04} and 0.556
fm$^{-1}$ fm$^{-1}$ \cite{Des94,Ang03}.  Moreover, our central
value for $C^2$ is  2.4$\sigma$ lower than that obtained in
\cite{Des04} for the MN potential.  It follows from here that the
update of the microscopic ($^3He+\alpha +p$) cluster calculation
\cite{Des04} performed for the MN potential correctly reproduces
the normalization of the tail of the radial overlap function of
$^8B$ nucleus in the ($p+^7Be$)-channel. As the microscopic
cluster-model calculations reproduce  rather well other observed
spectroscopic nuclear properties of $^8B$ \cite{Des94,Des04}, as
shown in the present work, the data obtained by us   for the ANC's
(NVC's) should also be included as an observable nuclear property
of $^8B$.

\subsection{Astrophysical S-factor for the  $^7Be(p,\gamma)^8B$
reaction at solar energies}

\begin{figure}[!h]
\includegraphics{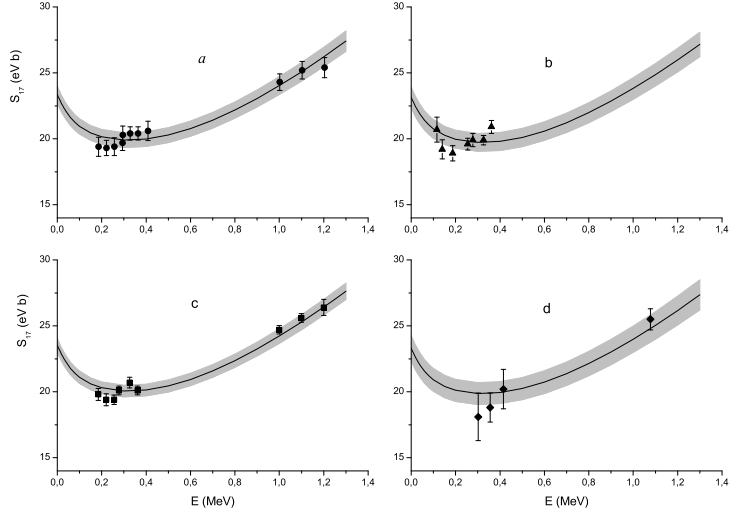} 
\caption{\label{fig5}The direct
astrophysical S-factors for the $^7Be(p,\gamma)^8B$ reaction. The
filled symbols are experimental data taken from  Refs.\cite{Jun03}
(BE1$(a)$, BE3 L$(b)$ and BE3 S $(c)$) and  \cite{Baby03} $(d)$.
The solid lines are the results of our calculation with the
standard values of $r_o$ =1.20 fm and $a$ =0.52 fm. The width of
the bands corresponds to the uncertainty associated with that of
both the ${\cal R}(E,C^{(sp)})$ function and the  values of the
ANC's  $C^2$  given in Table \ref{table2}.}
\end{figure}

\hspace{0.7cm}  The equation (\ref{18}) and the weighed means of
the ANC's obtained from the analysis of each of the experimental
astrophysical S-factors can be used for calculating the
corresponding astrophysical S-factor for the
 $^7Be(p,\gamma)^8B$  reaction   at solar energies. At
first, we tested again the  fulfilment of the condition (\ref{15})
in the same way as it is done above for  E$\ge$ 116 keV. Similar
results  plotted in Fig.\ref{fig1}  are also observed for a
dependence of the ${\cal {R}}(E,C^{(sp)})$ function on the single
- particle ANC, $C^{(sp)}$, at solar energies. The results of
extrapolation of the astrophysical S-factor for   values of $E<$
116 keV obtained by us are displayed in Figs.\ref{fig5} and
 in the fourth-sixth columns of Table \ref{table2}.
In these figures, filled circles and filled triangles, filled
squares and  filled diamonds correspond  to the experimental  data
from \cite{Jun03,Jun02} and \cite{Baby03}, respectively.

\begin{figure}[!h]
\includegraphics{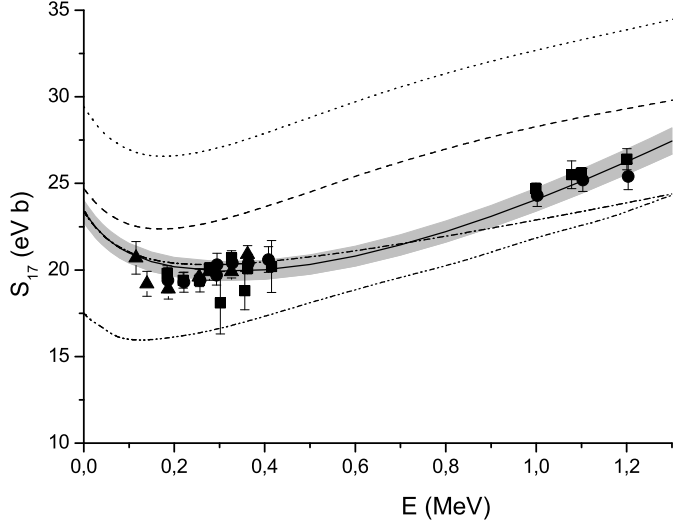} 
\caption{\label{fig6}The direct
astrophysical S-factor for the $^7Be(p,\gamma)^8B$ reaction. The
filled symbols, the solid line and the width of the band denote
the same as Fig.\ref{fig5}. The dashed (dotted) line is the
results taken from Ref.\cite{Des04} for the MN (V2)-potential for
the NN-potential. The dashed -dotted- dotted line is the result
taken from Ref.\cite{Xu94}. The dot-dashed lines are the result of
our calculation obtained from (\ref{23}) with ${\rm
S_{17}(0)}$=23.40 eVb.}
\end{figure}

In Fig.\ref{fig5}, the solid lines present    our calculations
performed with the standard  values of geometric parameters
$r_o$=1.20 fm and $a$=0.65 fm both for the bound ($p+^7Be$) state
and for $p^7Be$-scattering state using the corresponding weighted
means of the ANC - values from the second-fifth lines of the
second column of Table \ref{table2}. There the width of each band
for the curves corresponds to the a.s.e., which includes
uncertainties both for the corresponding ANC's and that in
$R(E,C^{(sp)})$. The results of extrapolation of the astrophysical
S-factors for third values of $E$ ($E$=0, 20 and 50 keV)
recommended by us are presented in the second-fifth lines of the
fourth-sixth columns of Table \ref{table2}. In Fig.\ref{fig6}, the
solid line presents our calculation performed also with the same
standard values of geometric parameters but with the weighted mean
of the recommended by us ANC's - value, $(C^{exp})^2$=0.628$\pm$
0.017 fm$^{-1}$. There the width of the band for the curve
corresponds to the a.s.e., which includes uncertainties both for
the  used ANC's and that in $R(E,C^{(sp)})$. As it is seen from
Figs.\ref{fig5} and \ref{fig6}, equation (\ref{18}) allows us to
perform a correct extrapolation of the corresponding astrophysical
S-factor at solar energies practically in an independent way when
the   ANC values for $p+^7Be\rightarrow ^8B$ are known. For
example, the ${\rm S_{17}(0)}$ value recommended by us is
$S_{17}(0)$=23.40$\pm$ 0.63 eVb.

For comparison, in Figs.\ref{fig6} the results of Ref.\cite{Des04}
(dashed and dotted lines ), obtained within the microscopic
cluster-model using  the MN- and V2-forms for the NN-potential,
respectively, are also presented. Besides, there the result of
Ref.\cite{Xu94} (dashed-dotted-dotted line), obtained within the
standard two-body potential model using the predicted values of
the ANC's (NVC)'s  for the M3YE-potential \cite{Mukh90} and the
value of the Barker spectroscopic factors being equal to 0.250 for
I=1 and 0.765 for I=2  for the two spin (I) configurations
($Z$=1.025) \cite{Bar80,Bar83}, is also plotted. The figure shows
strong overestimation (underestimation)in absolute values of the
calculated in Ref.\cite{Des04}(\cite{Xu94}(the dashed and dotted
lines (dashed- dotted-dotted line)) in respect to both the
experimental data, while our result (the solid line) reproduces
equally well both the energy dependence and the absolute values of
the experimental data.

The   astrophysical S-factor calculated by us and plotted in
Fig.\ref{fig6} by the solid line can be compared with the rational
expansion \cite{Jen98}
\begin{equation}
S_{17}(E)/S_{17}(0)=0.0408/(E+0.1375)+0.7033+0.2392E,
 \label{23}
\end{equation}
with the polynomial formulae \cite{Baye00}
\begin{equation}
S_{17}(E)= 23.604 -50.151E+515.802E^2
 \label{24}
\end{equation}
for the T potential and
\begin{equation}
S_{17}(E)=23.403-39.242E+330.094E^2 \label{25}
\end{equation}
 for the B potential. Here $S_{17}(E)$  is  in units of  eVb and $E$ in MeV.
One notes that the polynomial formulae  (\ref{24}) and (\ref{25})
were deduced by us from Eq.(43) of Ref.\cite{Baye00} and from  the
data presented in Table 4 there. A comparison  of  the result of
the present work  with that obtained from
Eq.(\ref{23})(Eqs.(\ref{24}) and (\ref{25})) shows that the ratio
$S_{17}(E)/S_{17}(0)$ ($S_{17}(E)$) obtained from formula
(\ref{23}) (formulae (\ref{24}) and (\ref{25})) changes from 0.98
to 1.10 in the energy range 0$\le$ E$\le$ 1200 keV (from 0.99 to
0.88 and 0.99 to 0.92, respectively, in the energy range 0$\le$
E$\le$ 100 keV) times with increase of E. It is seen that the
extrapolation formula proposed in \cite{Jen98}(\cite{Baye00}) in
the aforesaid energy range has    an accuracy not exceeding
$\sim$ 12$\%$ ($\sim$ 10$\%$) in respect to our results. As an
illustration, in Fig.\ref{fig6} the result of our calculation  for
$S_{17}(0)$ obtained from (\ref{23}) with ${\rm S_{17}(0)}$=23.40
eVb   is also displayed (dot-dashed lines). The figure shows that
at $E\gtrsim$ 1.0 MeV a noticeable discrepancy in absolute values
of the calculated $S_{17}(E)$ obtained in the present work
formulae and (\ref{23}) at the same normalization in the point E=0
($S_{17}(0)$=23.40 eVb ) occurs. Besides, as the energy E
increases (100 $<$ E$\le$ 400 keV), the discrepancy between the
result of the present work   and  that obtained from Eq.(\ref{24})
increases (up to 67\%).

Our result for $S_{17}(0)$ is about 3$\sigma$ larger and rather
larger than that of the extrapolation obtained in
Refs.\cite{Jun03,Baby03} and the deduction in \cite{Tab06},
respectively. Apparently, one of the possible reasons of the
observed discrepancy between our result and that of
Ref.\cite{Tab06} is connected with the underestimated value of
$C^2$ obtained in [33-35] in respect to our result. Such
conclusion can be explained by the fact that the ratio ($R_C$) of
the ANC $(C^{exp})^2$ of the present work to that recommended in
\cite{Tab06}($R_C$=1.35$\pm$ 0.15) practically coincides with the
ratio ($R_S$) for the corresponding $S_{17}(0)$ ($R_S$=1.29$\pm$
0.13). It follows from here that the obtained in [33-35] values of
ANC's for $p+^7{\rm Be}\to^8B$ may not have the necessary accuracy
for the astrophysical application because of their model
dependence \cite{AMukh3,Yar97,Igam072}.  From this point of view
one can understand the statement  of authors of
Refs.\cite{Jun03,Jun02,Baby03} on difficulties of determination of
all uncertainties related to the model dependence of the obtained
in [33-35] $S_{17}(0)$ values.

The resulting $S_{17}(0)$ obtained by us is in an excellent
agreement with the values $S_{17}(0)$=23.6 and 23.4 eVb
\cite{Baye00}, which were obtained within the standard two-body
potential model by using two (T and B) of sets of the parameters
for the Woods-Saxon potential, respectively, and the values of
the Barker spectroscopic factors \cite{Bar80,Bar83}.  In
Ref.\cite{Baye00}, the geometric parameters of the adopted
potential were taken as $r_o$=1.48 fm and $a$=0.52 fm
\cite{Tomb}(T potential ) and $r_o$=1.20 fm and $a$=0.65 fm
\cite{Bar80,Bar83}(B potential). Nevertheless,   in reality  the
model dependence of   the $S_{17}(0)$ values  calculated in
\cite{Baye00}  on the  used  in \cite{Baye00} values of the
spectroscopic factors $Z_I $ ($I$=1  and 2)   should be expected.
For  example, the  values of $S_{17}(0)$=27.4 and 27.2 eVb are
obtained for the T  and B potentials , respectively, as the
spectroscopic factors $Z_I $ are taken  equal to 0.3231 for I=1
and 0.8572 for I=2 ($Z$=1.1803) \cite{CK67,David03} instead of the
aforesaid ones used in \cite{Baye00}.

It should be mentioned here the refined  values of
$S_{17}(0)$=24.69 and 29.45 eVb \cite{Des04}, obtained for the MN-
and V2-forms of the NN-potential by updating of the microscopic
cluster calculation \cite{Des94}, respectively. It is seen from
here that  the value of $S_{17}(0)$ obtained in the present work
practically coincides within the experimental uncertainty with
that obtained in \cite{Des94} for the MN-form of the NN-potential.
Such matching does happen since  the ANC value obtained in the
present work and that calculated in \cite{Des94} are closer each
other. This confirms the conclusion made by authors of
Refs.[29-31,42] about the fact that an absolute value of
$S_{17}(0)$ is mainly determined by the two ANC's for
$p+^7Be\to^8B$ corresponding to the 1$p_{1/2}$ and 1$p_{3/2}$
proton orbitals in $^8B$. Besides, the recommended by us weighted
mean of $S_{17}(20 keV)$=22.8$\pm$ 0.6 eVb  at the energy close to
the Gamow peak is in a good agreement with $S_{17}(20 keV)$= 22.37
eVb obtained in \cite{Des04} for the MN-form of the NN-potential
but it differs noticeably on those of $S_{17}(20 keV)$=20.6$\pm$
0.5(expt) $\pm$ 0.6(theor)eVb recommended in \cite{Jun03} and
$S_{17}(20 keV)$=26.56 eVb \cite{Des04} obtained for the V2-form
of the NN-potential. Hence the slope of the solid curve in
Fig.\ref{fig6} near E$\sim$ 0 noticeably differs from that of the
dotted line obtained in \cite{Des04}   for the V2-form of the
NN-potential (see subsection 3.3). Therefore, uncertainties in
solar-model calculations would ultimately be reduced, if the value
of $S_{17}(20 keV)$ recommended by us should be used there.

We would like  to note the result   of $S_{17}(0)$=18.6$\pm$ 0.4
(expt)$\pm$ 1.1 (extrapolation) eVb obtained in Ref.\cite{David03}
from the analysis of the $^7Be$ longitudinal momentum
distributions performed within the potential model for  the
Coulomb breakup of $^8B$ via the $^{208}{\rm Pb}(^8{\rm
B},p^7Be)^{208}{\rm Pb}$ reaction. It is seen that the value of
$S_{17}(0)$ is also less than the present result. But, the method
used in Ref.\cite{David03} is also still subject to uncertainties
related to the model dependence of the extracted $S_{17}(0)$ value
in respect to both the spectroscopic factors (of 0.3231 for I=1
and 0.8572 for I=2 \cite{CK67}). This is related to the result of
Ref.\cite{Xu94}, $S_{17}(0)\approx$ 17.6 eVb, which was obtained
using the aforesaid Barker spectroscopic factors.

The   results of the Coulomb breakup [22-27,58]  give also the
underestimated values for $S_{17}(0)$ in respect to our result. In
this connection one should draw attention to the following. From
our point of view in the studies of authors of Refs.[22-27], which
were carried out within the semiclassical model [63-65] and its
modification \cite{Esben96}, the extracted values of the
astrophysical S-factors  are affected by various uncertainties
arising  because of higher-order effects \cite{Typel,Bert95},
including   the three-body Coulomb postdecay acceleration effects
(TBCPDAE) [32,69-72] in the final state of the breakup reactions
under consideration. It should be noted that  the TBCPDAE may also
play an important role in the pure  Coulomb breakup amplitude  at
relative distances of the colliding nuclei $R$ exceeding a value
of the radius $R_N$ but less than an impact parameter $b$ ( i.e.,
$ R_N < R_0 \le R < b$  in the region $r/R<<1$ , where $R_0$ is
the minimal distance the colliding nuclei approach at which the
nuclear interaction vanishes\footnote{According to \cite{Baur},
the value of $R_0$ is taken to be equal to $R_0=R_N +\pi Z_{Pb}Z_B
e^2/4E_i <b$ [63-65, 73], where $Z_je$ is a charge of the particle
$j$, $b$ is an impact parameter and  $E_i$ is the relative kinetic
energy of the colliding nuclei.} and $R_N $ is the nuclear
interaction radius between the colliding nuclei) \cite{Typel},
since the purely Coulomb breakup amplitude is not expressed
through the amplitudes of the photodisintegration process $\gamma
+^8B\to p+^7Be$ because the $p^7Be$-scattering wave function
depends on the local relative momentum of the $^7Be$ and proton
${\bf {q}}_{p^7Be}({\bf R})$, but not on the unaltered relative
momentum ${\bf k}$ [69-72,74,75]. The local relative momentum of
the $^7Be$ and proton in the final state may noticeably differ on
the unaltered relative momentum of $^7{\rm Be}$ and proton, i.e.,
when an influence of the Coulomb field of the multicharged
$^{208}{\rm Pb}$-ion in the final state of the Coulomb breakup
considered vanishes. As an example, we consider the Coulomb
breakup $^{208}{\rm Pb}(^8B,p^7Be)^{208}{\rm Pb}$ reaction at
energies of $^8B$ ion of 83 MeV/N \cite{David01a} and 254 MeV/N
\cite{Schum03} in which a value of the nuclear $ PbB$ interaction
radius ($R_N $ ) is taken as in Ref.\cite{Rolfs79}, that is $R_N
$=1.36(208$^{1/3}$+8$^{1/3}$)=10.78 fm. Then the calculated value
of the parameter $R_0$ is to be 11.51 and 11.02 fm for the
projectile energy of 83
 and 254 MeV/A, respectively. The kinematics of the
Coulomb breakup considered is chosen so that the relative momentum
(${\bf k}$) of the proton and $^7Be$  is parallel to the relative
momentum  of the center of mass of the ($p^7Be$)-pair and the
$^{208}{\rm Pb}$ nucleus (${\bf {k}}_f$). Then for the local
momentum ${\bf {q}}_{p^7Be}({\bf R})$ in the approximation up to
terms of the order
 $O(R^{-2})$  and in the limit ($\hat{\bf
k}_f\hat{\bf R}$)$\to$ 1\footnote{One notes that  the limit of
($\hat{\bf k}_f\hat{\bf R}$)$\to$ 1 for $\mid {\bf q_{p^7{\rm
Be}}(R)}\mid$ corresponds to a minimum value of $\mid {\bf
q_{p^7Be}(R)}\mid$ ($q(k,k_f;R)$) reached an inside of the range
$-1<(\hat{\bf k}_f\hat{\bf R})\le 1$ at a fixed value of $R$.},
similar as in Ref.\cite{Igam97a,Igam}, one has
\begin{equation}
\mid {\bf q_{p^7Be}(R)}\mid\ge q(k,k_f;R)\equiv k[1+\Delta
(k,k_f;R)]. \label{1i}
\end {equation}
Here $\Delta (k,k_f;R)=C( k_f)/kR$ and $\hat{\bf x}={\bf x}/x$ is
a unit vector, where $C( k_f)$=0.075$\eta_f(k_f)$ and
$\eta_f(k_f)=Z_{B}Z_{Pb}e^2\mu_{BPb}/k_f$. For illustration,
Table \ref{table3} shows the distinction between $k$ and
$q(k,k_f;R)$ within the momentum (energy) region of 0.0649$\le
k\le$ 0.2050 fm$^{-1}$ (0.10$\le E\le$ 1.00 MeV) for $R=R_0$=11.51
and 11.02 fm. As it is seen from Table \ref{table3}, the
calculated  value  of $q(k,k_f;R_0)$ differs noticeably from the
corresponding one of $k$, and a degree of this distinction is
enhanced with a decrease both of the value of $k$ (or E) and  of
the  $^8B$ projectile energy. But as the relative distance of the
colliding particles $R$ increases, the distinction between
$q(k,k_f;R)$ and $k$ becomes less since the influence of the
Coulomb field of $^{208}{\rm Pb}$-ion on the relative momentum of
the ($p^7Be$)-pair is also decreased because of a decrease of the
term of $C( k_f)/R$ in the r.h.s. of Eq.(\ref{1i}). It follows
from here that  for  each   fixed value of $R(= R_0)$, even the
minimal value of $\mid {\bf q_{p^7{\rm Be}}(R)}\mid$ calculated at
different values of $k$ may also differ noticeably from the
unaltered  corresponding relative momentum $k$. Therefore, a use
of the Coulomb breakup $^{208}{\rm Pb}(^8{\rm B},p^7Be)^{208}{\rm
Pb}$ cross sections for getting the information about the
astrophysical S-factors of the $ ^7{\rm Be}(p,\gamma)^8B$ reaction
for the energy region  of E$\lesssim$ 1.0 MeV eliminates the
TBCPDAE completely. This fact was not taken into account by
authors of Refs.[22-27]. Perhaps that is one of the possible
reasons of the aforesaid systematically observed discrepancy
between the results of the $S_{17}(E)$ obtained by the indirect
and direct ways. This conclusion is also confirmed by the results
of Refs.\cite{Alt03,Alt05} obtained from the analysis of the
double differential cross sections (DDCS) for the Coulomb breakup
$^{208}{\rm Pb}(^8B,p^7Be)^{208}{\rm Pb}$ reaction at projectile
energies 46.5 and 83 MeV/A performed on the basis of the
asymptotic three-body approach.  As shown in Ref.\cite{Alt05}, as
the   energy E decreases and the scattering angle  $\theta$ is
enhanced, the TBCPDAE become well important.

Thus,  the overall normalization of the direct astrophysical
S-factors at extremely low energies for the reactions under
consideration is mainly determined by the ANC - values for
$p+^7{\rm Be}\rightarrow ^8B$, which in turn can be determined
rather well from an analysis of the precisely measured
astrophysical S-factors \cite{Jun03,Jun02,Baby03} in a model
independent  way, and the found values of the ANC's allow one to
perform correct extrapolation of the astrophysical S-factors for
the direct radiative capture $^7Be(p,\gamma)^8B$   reaction at
solar energies, including $E=0$.

\subsection{The effective range parameters for
 $p^7Be$  scattering}

\hspace{0.6cm} Here it is of interest  to obtain  experimental
values of the $s$ wave average scattering length ${\bar
a}_0^{exp}$ \cite{Bay002} and the $p$ wave effective range
parameters (the scattering lengths $a_{1I}^{exp}$ and the
effective radius $r_{1I}^{exp}$, where $I$ = 1 and 2)
\cite{Igam07} using the expression for the logarithmic derivative
of $S_{17}(E)$ at the point of E=0
($s_1=S_{17}^{\prime}(0)/S_{17}(0)$) derived in \cite{Bay002} for
the B potential   and the values of ANC's for the $p+^7Be\to^8B$
obtained by us above. To this end one applies the formulae
(28)-(30) from \cite{Bay002}, which determine the dependence of
$s_1$ on the   $s$ wave average scattering length ${\bar a}_0$ for
the B potential, and the formulae (34) and (25) obtained in
\cite{Igam07}\footnote{See the foot-note 3 in the present work
too.}  for the relation between the ANC's and the scattering
lengths $a_{1I}$ and the effective radius $r_{1I}$.

From the astrophysical S-factor calculated by us   and the formula
$s_1\approx$ -1.81(1+0.087${\bar a}_0)$ \cite{Bay002} one obtains
$s_1^{exp}$=-1.699$\pm$ 0.066 MeV$^{-1}$ and ${\bar
a}_0^{exp}$=-7.070$\pm$ 0.091 fm. One notes that   the  polynomial
formula (\ref{25}) and the aforesaid expression for $s_1$
\cite{Bay002} give the values of $s_1$=-1.677 MeV$^{-1}$ and  the
$s$ wave average scattering length ${\bar a}_0$=-8.456 fm, which
are in  an  agreement with the result obtained in the present
work. The resulting the ${\bar a}_0^{exp}$ value obtained by us is
also in agreement with the value ${\bar a}_0\simeq$ =-7$\pm$ 3 fm
recommended in Ref.\cite{Ang03}.   Besides, one would like to note
the results of calculation of $s_1$ obtained in Ref.\cite{Bar06}
using the  values of $S(E)$ calculated in Ref .\cite{Des04}(
Ref.\cite{ Des94}) for MN and  V2 ( V2) forms for the
NN-potential. They are $s_1$=-1.86 MeV$^{-1}$ for MN and
$s_1$=-1.92 MeV$^{-1}$ for V2 \cite{ Bar06,Des04} as well as
$s_1$=-1.97 MeV$^{-1}$ for V2 \cite{ Bar06,Des94}. It is seen from
here that the first of them is in a good agreement with
$s_1^{exp}$ obtained in the present work. But two of the latter
those   differ noticeably on the aforesaid value of $s_1^{exp}$,
that is,  the slope of $S(E)$ near $E$=0 obtained in
Ref.\cite{Bar06} by using the results \cite{Des94,Des04} for the
V2-potential becomes slightly steeper than that defined in the
present work.

Using our result  for the weighted means of the experimental ANC's
and the following relations for the ANC's written in another
coupling modes  $(C^{exp}_{I=2})^2
=(C^{exp}_{p_{1/2}}+C^{exp}_{p_{3/2}})^2/2$ and $(C^{exp}_{I=1})^2
=(C^{exp}_{p_{1/2}}-C^{exp}_{p_{3/2}})^2/2$ \cite{Blok77}, from
the formulae (34) and (25) of Ref.\cite{Igam07} one obtains the
following $p$ wave effective range parameters
$r_{12}^{exp}$=-0.177$\pm$ 0.001 fm$^{-1}$ and
$a_{12}^{exp}$=(1.374$\pm$ 0.635)x$10^4$ fm$^3$ for $l_i$=1 and
$I$=2 and $r_{11}^{exp}$=-0.806$\pm$ 0.107 fm$^{-1}$ and
$a_{11}^{exp}$=(5.296 $\pm$ 0.867)x$10^2$ fm$^3$ for $l_ i$=1 and
$I$=1.

The results given here for the $s$ wave average scattering length
(${\bar a}_0^{exp}$) and the $p$ wave effective range parameters
($r_{1I}^{exp}$ and $a_{1I}^{exp}$) can be considered as
``indirect measurements'', of these parameters for $s$ and $p$
waves $p^7Be$ scattering since they are also determined by the
``indirect measured'', values of the $s_1$ and the ANC's.

\section{Conclusion}

\hspace{0.6 cm} The modified two-body potential approach proposed
recently in Ref.\cite{Igam07} for  the peripheral direct radiative
capture $A(a,\gamma)B$ reaction of astrophysical interest is
applied to    the reanalysis of the experimental astrophysical
S-factors, $S^{exp}_{17}(E)$, for the $^7Be(p,\gamma)^8B$
precisely measured recently  at energies 116$\le$ E$\le$ 400 keV
and 1000$\lesssim$ E$\lesssim$ 1200 keV\cite{Jun03,Jun02,Baby03}.

  It is demonstrated that the
experimental astrophysical S-factors of the reaction  under
consideration measured in the energy regions  E$\le$ 0.4 MeV  and
1000$\lesssim$ E$\lesssim$ 1200 keV can be used as an independent
source of getting the information about the ANC's (or NVC's) for
$p+^7Be\rightarrow ^8B$.    The weighed means of the ANC's (NVC's)
for the $p+^7Be\rightarrow ^8B$ are obtained. They have to be
$(C^{exp})^2= (C^{exp}_{p_{1/2}})^2
+(C^{exp}_{p_{3/2}})^2$=0.628$\pm$ 0.017 fm$^{-1}$ ($\mid
G\mid_{exp}^2$=0.114$\pm$ 0.003 fm ),
$(C^{exp}_{p_{3/2}})^2$=0.558$\pm$ 0.015 fm$^{-1}$ ($ \mid
G_{p_{3/2}}\mid_{exp}^2$=0.0995$\pm$ 0.0027 fm) and
$(C^{exp}_{p_{1/2}})^2$=0.070$\pm$ 0.002  fm$^{-1}$
 ($\mid G_{p_{1/2}}\mid_{exp}^2$=0.0145$\pm$ 0.0004 fm).
The uncertainty in the ANC  (NVC )-values  obtained by us includes
the experimental errors for the experimental astrophysical
S-factors, $S^{exp}_{17}(E)$ and that of the used approach.

Our result for $(C^{exp})^2$ (or $|G |_{exp}^2$) differs
noticeably ( up to 35\%) from those obtained by authors of
Refs.[33-35] and \cite{Trache01} from the analysis of the
peripheral proton transfer reactions and the $^8B$ breakup
reactions, respectively,  and is in an   agreement with that
calculated by author of \cite{Des04} within the microscopic
cluster method for the MN-form  for the NN-potential.

The obtained value  of the ANC  was used for   an extrapolation of
the astrophysical S-factor  of the reaction under consideration at
energies less than 116 keV, including $E=0$.  In particular, for
the astrophysical S-factor $S_{17}(0)$ the value of ${\rm
S_{17}(0)}$=23.40$\pm$ 0.63 eVb has been obtained. Our result for
$S_{17}(0)$ is a bit over 2$\sigma$ larger than that obtained in
\cite{Des04} and noticeably more than  the results of ${\rm
S_{17}(0)}$=21.4$\pm$ 0.5(expt)$\pm$ 0.6(theor)   and 21.2$\pm$
0.7 eVb recommended in Refs.\cite{Jun03} and \cite{Baby03},
respectively. Besides,  the  $S_{17}(0)$ value obtained in the
present work differs noticeably from that recommended in
Refs.[33-35] and Ref.[22-27,52] deduced within the ANC technique
and the Coulomb breakup reaction calculations, respectively.  The
obtained values of $S_{17}(0)$, $S_{17}^{\prime}(0)$ and ANC's
were used for the estimation of values of the $s$ wave average
scattering length and the $p$ waves effective range parameters.

\newpage
\section{Acknowledgments}

The authors are deeply grateful to  S.V. Artemov and  R.J.
 Peterson for useful  discussions. The work has been supported by the Forum on International Physics
 of the American Physical Society (2006) and
by   The Academy of Science  of The Republic of  Uzbekistan (Grant
No.FA-F2-F077).  One of authors (R.Y.)  thanks  Departments of
Physics of the University of Colorado (Boulder) and the Colorado
 School of Mines  (Golden) for  hospitality ( August 2006), where the work has partially  been performed.

\newpage
\newpage
\newpage
\newpage
\newpage
\newpage
\newpage
\landscape
\begin{table}[h]
\caption{\label{table1}The dependence of the single-particle
astrophysical S - factor $\tilde{S}(E)$, the function ${\cal
R}(E,C^{(sp)})$, ANC's ($C^2$)  and spectroscopic factors ($Z$) on
the single-particle ANC $C^{(sp)}$.}
\begin{tabular}{|c|c|c|c|c|c|c|c|c|c|c|c|c|c|c|}\hline
$r_o$&$a$&$C^{(sp)}$&\multicolumn{3}{c|}{$C^2$}&\multicolumn{3}{c|}
{$Z$}&\multicolumn{3}{c|}
{$\tilde{S}$} &\multicolumn{3}{c|}{$R(E,C^{sp})$}\\
(fm)&(fm)& (fm$^{-1/2})$&
\multicolumn{3}{c|}{(fm$^{-1})$}&\multicolumn{3}{c|}{}
&\multicolumn{3}{c|}{(eVb)}&\multicolumn{3}{c|} {(eV b fm)} \\
\hline
0.90&0.52&0.613&0.623&0.642&0.603&1.657&1.707&1.604&12.495&12.072&15.837&32.080&32.083&42.090\\
\hline
1.08&0.56&0.684&0.623&0.642&0.603&1.333&1.373&1.289&15.530&15.002&19.706&33.206&32.078&42.136\\
\hline
1.20&0.65&0.768&0.625&0.645&0.606&1.059&1.094&1.028&19.545&18.829&24.715&33.145&31.931&41.914
\\ \hline
1.32&0.70&0.831&0.626&0.648&0.609&0.907&0.926&0.883&22.828&21.949&28.918&33.087&31.814&41.709\\
\hline
1.50&0.76&0.920&0.628&0.653&0.614&0.741&0.771&0.725&27.942&26.730&35.050&32.984&31.553&41.375
\\ \hline
\multicolumn{3}{|c|}{$E$(MeV)$\rightarrow$}&0.1156&0.4081&1.2032&0.1156&0.4081&1.2032&0.1156&0.4081
&1.2032&0.1156&0.4081&1.2032\\ \hline
\end{tabular}
\end{table}
\endlandscape
\newpage
\begin{table}[h]
\caption{\label{table2}The weighted means of the ANC - values
($C^{exp})^2$ for   $p +^7Be\rightarrow ^8B$, NVC's $\mid
G\mid^2_{exp}$ and the calculated values of $S_{17}(E)$ at
energies $E$=0, 20 and 50 keV .}
\begin{tabular}{|l|c|c|c|c|c|}\hline
Exp.&$(C^{exp})^2$ (fm$^{-1})$&$\mid G\mid^2_{exp}$
(fm)&S(0)keV)(eVb) &S(20 keV) (eVb)&S(50 keV) (eVb)\\ \hline
\cite{Jun03,Jun02}(BE1)&0.625$\pm $0.016&
0.114$\pm$0.003&23.354$\pm$0.647&22.645$\pm$0.627&21.815$\pm$0.605
\\ \hline
\cite{Jun03,Jun02}(BE3L)&0.621$\pm$0.021&0.113$\pm$0.004&23.144$\pm$0.789&22.441$\pm$0.765&21.619$\pm$0.737
\\ \hline
\cite{Jun03,Jun02}(BE3S)&0.631$\pm$0.014&0.115$\pm$0.003&23.537$\pm$0.519&22.821$\pm$0.503&21.986$\pm$0.485
\\ \hline
\cite{Baby03}&0.625$\pm$0.026&0.114$\pm$0.005&23.309$\pm$0.967&22.600$\pm$0.938&21.773$\pm$0.903
\\ \hline
\end{tabular}
\end{table}

\begin{table}[h]
\caption{\label{table3} The dependence of  the calculated minimal
values of the relative local momentum ($q=q(k,k_f,R_0)$) for the
($p^7Be$)-pair formed in  the $^{208}{\rm Pb}(^8B,p^7{\rm
Be})^{208}{\rm Pb}$ reaction for two projectile energies
($E_{^8{\rm B}}$) at different energies E ( or
   unaltered   relative momentums $k$.)}
\begin{tabular}{|c|c|c|c|c|c|}\hline
$E$ (MeV)&$k$ (fm$^{-1}$)&\multicolumn{2}{c|}{$E_{^8B}=83$
(MeV/A), $R_0$=11.51 fm}
  &\multicolumn{2}{c|}{$E_{^8B}=254$ (MeV/A), $R_0$=11.02 fm}\\
\cline{3-6}
 &&$q$ (fm$^{-1}$)&$q/k$&$q$ (fm$^{-1}$)&$q/k$ \\ \hline
 0.10&0.0649&0.111&1.71&0.0926&1.43 \\ \hline
 0.15&0.0795&0.126&1.58&0.107&1.35 \\ \hline
 0.20&0.0918&0.138&1.50&0.120&1.30 \\ \hline
 0.25&0.103&0.149&1.45&0.130&1.27 \\ \hline
 0.30&0.112&0.159&1.41&0.140&1.25 \\ \hline
 0.40&0.130&0.176&1.36&0.158&1.21 \\ \hline
 0.75&0.177&0.224&1.26&0.206&1.16 \\ \hline
 1.00&0.205&0.252&1.23&0.233&1.13 \\ \hline
\end{tabular}
\end{table}
\end{document}